\documentstyle[twocolumn,epsf,prl,aps]{revtex}

\begin{document}

\draft
\twocolumn[\hsize\textwidth\columnwidth\hsize\csname@twocolumnfalse%
\endcsname
\title{
Quasiparticle Density of States of Clean and Dirty {\it s}-Wave 
Superconductors \\ in the Vortex State}
\author{ 
M. Nohara$^{1,2}$, M. Isshiki$^{1}$, F. Sakai$^{1}$, and H. Takagi$^{1,2}$}
\address{
$^{1}$Institute for Solid State Physics, University of Tokyo, 
7-22-1, Roppongi, Minato-ku, Tokyo 106-8666, Japan \\
$^{2}$CREST, Japan Science and Technology Corporation, Japan}
\date{January 14, 1999}
\maketitle
\begin{abstract}
The quasiparticle density of states (DOS) in the vortex state has been 
probed by specific heat measurements under magnetic fields ($H$) for clean 
and dirty {\it s}-wave superconductors, Y(Ni$_{1-x}$Pt$_{x}$)$_{2}$B$_{2}$C 
and Nb$_{1-x}$Ta$_{x}$Se$_{2}$. 
We find that the quasiparticle DOS per vortex is appreciably 
$H$-dependent in the clean-limit superconductors, while it is 
$H$-independent in the dirty superconductors as expected from a 
conventional rigid normal electron core picture. 
We discuss possible origins for our observations in terms of the 
shrinking of the vortex core radius with increasing $H$. 
\end{abstract}
]

Determination of the electronic structure of magnetic vortex lines in 
type-II superconductors has been an issue of long standing interest.
Recent attention has focussed on its relation with the symmetry of the 
Cooper pairs:
For $s$-wave superconductors with an isotropic gap, there exist
low-lying bound quasiparticles around the center of the vortex lines, 
where the superconducting gap vanishes, 
and the vortex line has been simply viewed as a ``rigid'' normal electron 
cylinder with a radius of the coherence length $\xi$.~\cite{Caroli} 
On the other hand, for $d$-wave superconductors, 
quasiparticles are delocalized and spread outside the vortex cores 
because of the presence of the gapless region on the Fermi surface.
This difference manifests itself in, for example, a magnetic field 
($H$) dependence of the quasiparticle density of states (DOS) at the 
Fermi level $N_{0}(H)$; $N_{0}(H) \propto H$ for $s$-wave 
superconductors, while $N_{0}(H) \propto \sqrt{H}$ for $d$-wave 
superconductors with line nodes.~\cite{Volovik} 
However, even in $s$-wave superconductors, 
the following unsolved issues may potentially alter the above 
simple picture of the vortex lines: Shrinking of the core radius with 
decreasing temperature ($T$), due to the thermal population of the quantized 
levels for the bound excitations, 
has been predicted theoretically by Pesch and Kramer.~\cite{Kramer}
Shrinking of the core radius with increasing $H$ has been proposed 
based on scanning tunneling spectroscopy (STS)~\cite{Golubov} 
and muon spin rotation ($\mu$SR)~\cite{Sonier} measurements, 
though the physics behind this has not been fully explored yet. 
The de Haas - van Alphen (dHvA) effect has been observed in the 
mixed state where the cyclotron radius is much larger than 
$\xi$.~\cite{Graebner,Terashima}
This suggests the existence of delocalized quasiparticles 
even in the $s$-wave superconductors under high magnetic fields. 

In this letter, we report the low-temperature specific heat of 
pure and alloyed Y(Ni$_{1-x}$Pt$_{x}$)$_{2}$B$_{2}$C and 
Nb$_{1-x}$Ta$_{x}$Se$_{2}$ single crystals 
under magnetic fields, focusing on the $H$ dependence of the 
quasiparticle DOS. 
We will show that the $H$ dependence of the quasiparticle DOS at the 
Fermi level $N_{0}(H)$ is significantly influenced by the presence of 
disorder: While the DOS shows the $N_{0}(H) \propto H$ behavior in the 
alloyed samples, $N_{0}(H)$ shows significant deviation from the 
$N_{0}(H) \propto H$ behavior in the pure samples. 
We will discuss possible mechanisms of the deviations from the $H$-linear 
behavior and their relation with quasiparticle scattering.

Single crystals of Y(Ni$_{1-x}$Pt$_{x}$)$_{2}$B$_{2}$C were grown by a 
floating zone 
method,~\cite{S1} while iodine-assisted chemical transport  
was employed for the growth of Nb$_{1-x}$Ta$_{x}$Se$_{2}$.~\cite{S2}
Table I lists the superconducting and normal state parameters for 
the samples used in this study. 
As evident from the table, the pure samples, YNi$_{2}$B$_{2}$C and NbSe$_{2}$, 
are in the clean limit ($\xi \ll l$, where $l$ is the mean free path), 
while the alloyed samples, Y(Ni$_{0.8}$Pt$_{0.2}$)$_{2}$B$_{2}$C and 
Nb$_{0.8}$Ta$_{0.2}$Se$_{2}$, are in the intermediate 
range near the dirty limit ($\xi \geq l$).
To determine $N_{0}(H)$ in magnetic fields, we have measured 
low-temperature specific heat ($C$) from 1.4 K to 20 K under magnetic 
fields up to 12 T using a thermal relaxation 
calorimeter.~\cite{nohara}
The relative resolution of the measurements was better than 0.5 \%, 
and the absolute accuracy determined from the measurement of 
a Cu standard was better than 1 \%.

Specific heat data for Y(Ni$_{1-x}$Pt$_{x}$)$_{2}$B$_{2}$C under various 
magnetic fields 
parallel to the $c$ axis are shown in Fig.\ 1, plotted as $C/T$ versus 
$T^{2}$ for two samples (a) $x$ = 0.0 and (b) $x$ = 0.2. 
The measurements were performed with increasing temperature after 
field cooling from a temperature well above the superconducting 
transition temperature $T_{\rm c}$. 
For the pure sample, a clear jump is seen in the zero-field data 
at $T_{\rm c}$ = 15.4 K. 
The upper critical field $H_{\rm c2}(T)$ was determined from the specific heat 
data for a fixed $H$. The value at $T$ = 0, $H_{\rm c2}(0)$, was estimated by 
fitting the low-temperature data below 5 K to 
$H_{\rm c2}(T) = H_{\rm c2}(0)[1 - h(T/T_{\rm c})^{2}]$, 
where $h$ is a constant, giving an estimate of $H_{\rm c2}(0)$ = 8.0 T. 
This is consistent with a reported value of $H_{\rm c2}(0)$ = 8.1 T.~\cite{Y1}
The least squares fit of the 9 T data, which represent the normal- 
state specific heat, to $C = \gamma_{\rm N}T + \beta T^{3}$
between 1.4 K and 5 K provides a Sommerfeld constant $\gamma_{\rm N}$ = 
20.6 mJ/K$^{2}$mol. 
These values agree reasonably with those reported 
previously.~\cite{Michor}
For the $x$ = 0.2 sample, we obtain $T_{\rm c}$ = 12.1 K, 
$\gamma_{\rm N}$ = 14.7 mJ/K$^{2}$mol, and $H_{\rm c2}(0)$ = 4.3 T. 

Under magnetic fields, a finite amount of $T$-linear term, 
$\gamma(H)T$, appears in the specific heat well below $T_{\rm c}$ 
as shown in Fig.\ 1. 
The $\gamma(H)$ values at each field were estimated by extrapolating 
the $C/T$ versus $T^{2}$ data between 1.4 and 2.5 K linearly to $T$ = 0. 
The zero-field value $\gamma(0)$ for the pure sample 
is negligibly small, indicative of the high quality of the sample. 
A small residual $\gamma(0)$ of about 0.4 mJ/K$^{2}$mol is seen 
for the $x$ = 0.2 sample, which very likely originates from a 
tiny amount of normal-state secondary phase. 
In order to see the magnetic field dependence, $\gamma(H)/\gamma_{\rm N}$ is 
plotted in Fig.\ 3(a) as a function of $H/H_{\rm c2}$. 
A substantial difference between the two samples is evident. 
For the pure sample, $\gamma(H)$ first shows a marked increase 
at low fields, then gradually approaches the normal state value. 
Notably, $\gamma(H)$ reaches a value almost half that of $\gamma_{\rm N}$ at 
1 T ($H/H_{\rm c2} \simeq 0.13$). 
We observed essentially the same $H$-dependence of $\gamma(H)$ in the field 
direction $H \perp c$. 
Analogous nonlinear dependence of $\gamma(H)$ was also observed 
for LuNi$_{2}$B$_{2}$C.~\cite{nohara}
In contrast, for the $x$ = 0.2 sample, $\gamma(H)$ increases linearly 
with $H$ as $\gamma(H) = \gamma(0) + \gamma_{\rm N}H/H_{\rm c2}$
over a wide field range below $H_{\rm c2}$. 
$\gamma(H)$ does not show any field dependence above $H_{\rm c2}$, 
indicating that this region is indeed in the normal state.

Qualitatively similar behavior for $\gamma(H)$ was observed in 
Nb$_{1-x}$Ta$_{x}$Se$_{2}$.
Shown in Fig.\ 2 are the specific heat data for Nb$_{1-x}$Ta$_{x}$Se$_{2}$ 
under magnetic fields applied parallel to the $c$ axis. 
For the pure sample, $x$ = 0.0, we obtained $T_{\rm c}$ = 7.3 K, 
$\gamma_{\rm N}$ = 19.3 mJ/K$^{2}$mol, 
and $H_{\rm c2}(0)$ = 4.4 T. 
For the alloyed sample, $x$ = 0.2, we obtained 
$T_{\rm c}$ = 5.1 K, 
$\gamma_{\rm N}$ = 15.1 mJ/K$^{2}$mol, and $H_{\rm c2}(0)$ = 3.2 T. 
The magnetic field dependence of $\gamma(H)$ was estimated in the 
same manner as mentioned above, and is displayed in Fig.\ 3(b). 
A nonlinear behavior of $\gamma(H)$ is noticeable for the pure NbSe$_{2}$ 
sample, though not as significant as that observed in YNi$_{2}$B$_{2}$C. 
Similar nonlinear behavior of $\gamma(H)$ in the pure sample was 
reported by Sanchez {\it et al}.~\cite{Sanchez,Junod} 
In contrast, for the alloyed sample, Nb$_{0.8}$Ta$_{0.2}$Se$_{2}$, $\gamma(H)$
exhibits a linear dependence on $H$ below $H_{\rm c2}$ as observed in 
Y(Ni$_{0.8}$Pt$_{0.2}$)$_{2}$B$_{2}$C.

Since the number of vortices is scaled by $H$, the nonlinear 
behavior of $\gamma(H)$ in the pure superconductors implies that 
the DOS per single vortex depends on $H$ when the sample is clean. 
This contradicts the conventional picture that 
the vortex line should be viewed as a ``rigid'' normal electron cylinder, 
where the bound quasiparticles give rise to a $H$-independent DOS 
proportional to $N_{0}\pi\xi^{2}$ per vortex line, where 
$N_{0}$ is the DOS in the normal state.
However, once the electron mean free path $l$ is substantially reduced 
due to strong impurity scattering and 
$l \leq \xi <$ intervortex distance $R$, 
the conventional picture appears to hold true. 
We point out that the unusual $\gamma(H)$ behavior and the strong 
influence of disorder are universally observed for $s$-wave 
superconductors, although the detailed $\gamma(H)$ data are available 
only for a limited number of compounds: 
A nonlinear dependence of $\gamma(H)$ was reported on a clean 
{\it s}-wave superconductor CeRu$_{2}$,~\cite{Hedo} where the 
deviation is compatible with that observed in NbSe$_{2}$. 
By contrast, a linear behavior was reported
on a dirty superconductor Nb$_{77}$Zr$_{23}$.~\cite{Junod}
Both $H$-linear and nonlinear $\gamma(H)$ were observed on 
V$_{3}$Si,~\cite{Art,Stewart} which is very likely due to different 
degrees of disorder in the samples. 

We suggest that the apparent breakdown of the ``rigid'' normal 
electron picture originates from the $H$ dependent core radius. 
In the conventional picture, the vortex core is assumed to be rigid. 
However, if the core radius is $H$ dependent, the quasiparticle DOS 
per vortex should also be $H$ dependent, resulting in a nonlinear 
$\gamma(H)$. 
Indeed, direct measurement of the local DOS in NbSe$_{2}$ by STS gives 
evidence for the shrinking of the core radius.~\cite{Golubov}
Using the experimentally observed $\gamma_{\rm N}$ and $\gamma(H)$, 
we estimate the effective core radius $\rho_{\rm V}$ by defining 
$2\pi{\rho_{\rm V}(H)}^{2}\gamma_{\rm N} = \gamma(H)/(H/\Phi_{0})$, 
where $\Phi_{0}$ is the flux quantum. 
Note that $\rho_{\rm V}(H_{\rm c2}) = \xi(0)$. 
The variation of $\rho_{\rm V}(H)$ estimated in this way is shown in 
the inset to Fig.\ 3, 
and compared with those determined by STS for NbSe$_{2}$.~\cite{Golubov}
The $H$ dependence of $\rho_{\rm V}$ from these two independent probes 
appears to agree reasonably with each other. The difference in the 
magnitudes should not be taken as significant since there is an 
arbitrariness of a factor in the definition of $\rho_{\rm V}$. 

The shrinking of the core radius may be supported by an anomalous 
temperature dependence of $H_{\rm c2}(T)$ near $T_{\rm c}$. 
In Fig.\ 4(a), $H_{\rm c2}(T)$ curves deduced from the specific heat data 
are shown for the pure YNi$_{2}$B$_{2}$C where the deviation of $\gamma(H)$ 
from the linear behavior is most significant. 
A pronounced positive curvature is clearly seen near $T_{\rm c}$. 
This may be understood in terms of the shrinking of $\rho_{\rm V}$ 
(and hence $\xi$), which would lead to an enhancement of $H_{\rm c2}(T)$ 
with increasing $H$ through the relation 
$H_{\rm c2} = \Phi_{0} / 2 \pi \xi^{2}$. 
Indeed, from the suppressed initial slope at $T_{\rm c}$, 
${\rm d}H_{\rm c2}/{\rm d}T{\mid}_{T_{\rm c}} = - 0.30$ T/K, 
we can obtain an estimate for $\xi(0) \simeq 115$ {\AA} 
using $\xi(0) = 0.54[-\Phi_{0}/T_{\rm c}
({\rm d}H_{\rm c2}/{\rm d}T{\mid}_{T_{\rm c}})]^{1/2}$,
which is substantially larger than $\xi(0)$ = 65 {\AA}, but 
close to the estimate of the enhanced $\rho_{\rm V} \simeq 130$ {\AA} from 
$\gamma(H)$ at low magnetic fields limit (Fig.\ 3). 
Furthermore, as clearly seen in Fig.\ 4(a), the alloyed sample 
Y(Ni$_{1-x}$Pt$_{x}$)$_{2}$B$_{2}$C does not show any noticeable positive 
curvature, consistent 
with the $H$-independent $\rho_{\rm V}$.~\cite{Shulga}
A small positive curvature of $H_{\rm c2}(T)$ near $T_{\rm c}$ can be 
seen in the pure NbSe$_{2}$ as shown in Fig. 5(b) 
(See also ref.~\cite{Sanchez}) and CeRu$_{2}$,~\cite{Hedo} which is 
consistent with the observation of a definite but less pronounced deviation 
from the $H$-linear $\gamma(H)$ in these materials. 

By introducing a certain (repulsive) vortex-vortex interaction in the 
vortex lattice states, it was shown theoretically that the core 
radius indeed shrinks with increasing $H$.~\cite{Golubov,Sonier,Ichioka}
The physical origin of the vortex interaction is not clear yet. 
We speculate that this interaction is mediated by a coherent 
transfer of quasiparticles between the cores. 
The observation of dHvA oscillations in the mixed state of clean 
YNi$_{2}$B$_{2}$C~\cite{Terashima} and NbSe$_{2}$,~\cite{Graebner}
where the cyclotron radius is by far larger than 
the core radius, clearly indicates the presence of delocalized and 
highly coherent quasiparticles outside the vortex cores. 
These delocalized quasiparticles may give rise to a certain 
vortex-vortex interaction and result in a shrinking of the core 
radius with increasing $H$. Then the critical sensitivity of the 
$\gamma(H)$ behavior can be naturally understood, since the coherent 
motion of these quasiparticles will be substantially suppressed by the 
impurity scatterings, when  $l \sim \xi < R$. 
One may argue that the DOS associated with the delocalized 
quasiparticle itself may be the origin of nonlinear $\gamma(H)$ 
behavior. 
Hedo {\it et al}. recently proposed that the nonlinear behavior of 
$\gamma(H)$ in CeRu$_{2}$ originates from the contribution of delocalized 
quasiparticles with momentum $k$ perpendicular to $H$.~\cite{Hedo} 
The discussion is based on the theoretical study by Brandt 
{\it et al.},~\cite{Brandt} who showed that near $H_{\rm c2}$, 
while excitations with $k$ parallel to $H$ have a gap, 
those in planes perpendicular to $H$ are gapless due to the formation 
of the vortex line lattice. 
However, the theory by Brandt can be applicable only near $H_{\rm c2}$, 
and another theoretical approach with emphasis on the low field regime 
indicates that the local DOS outside the core is almost zero when 
$H/H_{\rm c2} < 0.4$.~\cite{Ichioka1}

Finally, we would like to comment on the pronounced nonlinear 
$\gamma(H)$ behavior in pure YNi$_{2}$B$_{2}$C compared with that of NbSe$_{2}$. 
We suspect that the presence of substantial gap anisotropy plays a role 
in YNi$_{2}$B$_{2}$C. 
The presence of a reduced gap region is suggested by the 
dHvA measurement.~\cite{Terashima} 
As seen in Fig.\ 1(a), the zero-field specific heat of pure YNi$_{2}$B$_{2}$C 
shows $T^3$ behavior rather than a thermally activated behavior 
down to low temperatures. 
This also supports the presence of a substantially small gap region 
on the Fermi surface, although $d$- or $p$-wave pairing is  
unlikely in YNi$_{2}$B$_{2}$C because of the insensitivity of the 
superconductivity to nonmagnetic Pt substitution. 
The reduced gap region on the Fermi surface may enhance 
quasiparticle transfer between the vortices and may result in the 
pronounced nonlinear $\gamma(H)$ in terms of the above speculative 
picture. 
In NbSe$_{2}$, zero-field specific heat exhibits thermally activated 
behavior at low temperatures, indicative the presence of a relatively 
isotropic gap. This is consistent with the smaller deviation from the 
linear behavior of $\gamma(H)$ and positive curvature of $H_{\rm c2}(T)$ in 
this compound. 

In summary, we have measured the specific heat of 
Y(Ni$_{1-x}$Pt$_{x}$)$_{2}$B$_{2}$C and Nb$_{1-x}$Ta$_{x}$Se$_{2}$ 
in the vortex state. Our results show that, in clean superconductors, 
the coefficient of the $T$-linear term in the specific heat, $\gamma(H)$, 
shows nonlinear behavior with $H$, suggesting that the DOS 
per vortex is $H$ dependent, while in dirty superconductors, 
$\gamma(H)$ is linear, suggesting a constant DOS per vortex. 
We suggest that these results may be understood in terms of the 
shrinking of the vortex core radius with increasing magnetic field. 
To strengthen this scenario, it is critically important to examine 
the $H$ dependence of $\rho_{\rm V}$ in dirty superconductors by direct 
probes such as STS and $\mu$SR. Such attempts are now in progress. 

We would like to thank K. Machida and M. Ichioka for valuable 
discussions. We acknowledge the critical reading of the manuscript by 
N. E. Hussey. 
This work was supported by a grant-in-aid of the Ministry of Education, 
Science, Sports, and Culture of Japan.

%-------------------------------Fig. 1---------------------------------
\begin{figure}
\begin{center}
\leavevmode
\epsfxsize=210pt    % set the width
\epsfbox{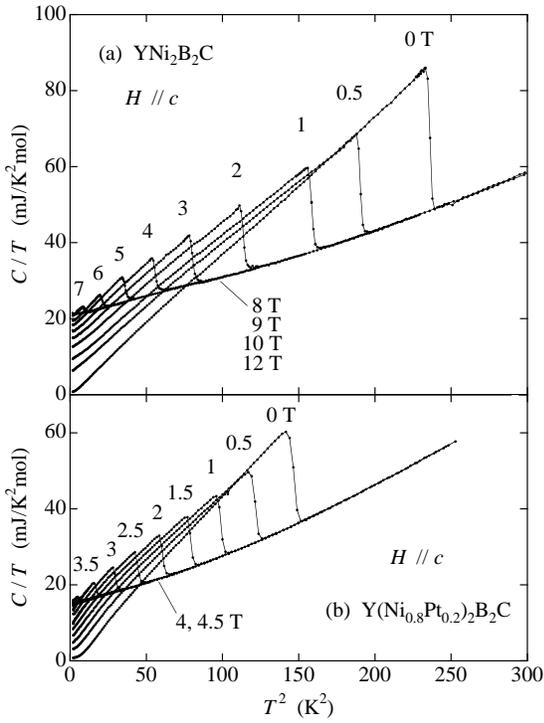}
\end{center}
\caption{
Specific heat divided by temperature $C/T$ for (a) YNi$_{2}$B$_{2}$C and (b) 
Y(Ni$_{0.8}$Pt$_{0.2}$)$_{2}$B$_{2}$C, as a function of $T^{2}$ under various 
magnetic fields. 
The solid lines are guides to the eye. 
}
\label{Fig1}
\end{figure}
%----------------------------------------------------------------------

%-------------------------------Fig. 2---------------------------------
\begin{figure}
\begin{center}
\leavevmode
\epsfxsize=210pt    % set the width
\epsfbox{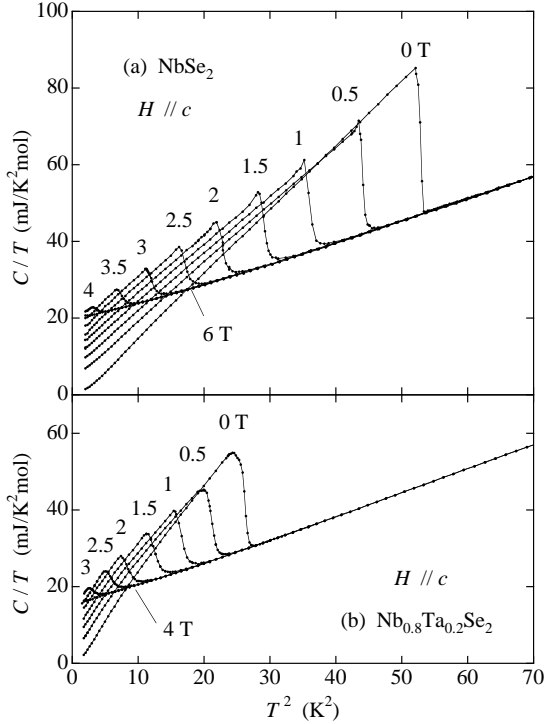}
\end{center}
\caption{
Specific heat divided by temperature $C/T$ for (a) NbSe$_{2}$ and (b) 
Nb$_{0.8}$Ta$_{0.2}$Se$_{2}$, as a function of $T^{2}$ under various magnetic 
fields. The solid lines are guides to the eye. 
}
\label{Fig2}
\end{figure}
%----------------------------------------------------------------------

%-------------------------------Fig. 3---------------------------------
\begin{figure}
\begin{center}
\leavevmode
\epsfxsize=210pt    % set the width
\epsfbox{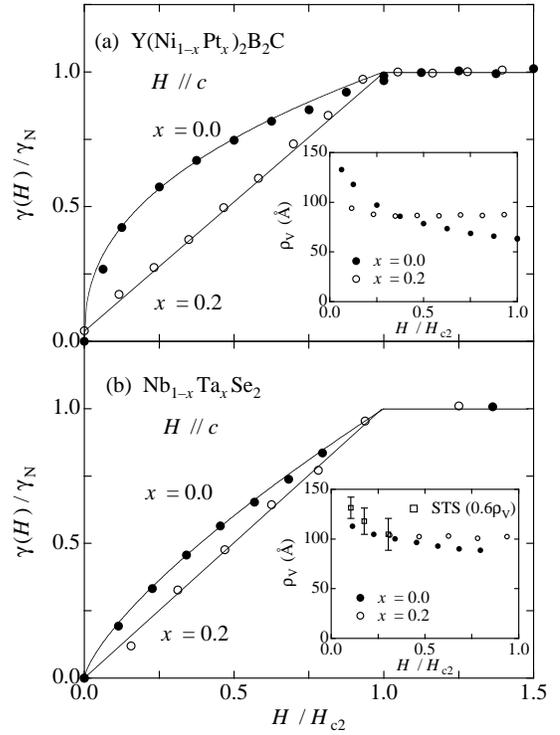}
\end{center}
\caption{
The coefficient of $T$-linear specific heat at low temperatures 
normalized by the Sommerfeld constant, $\gamma(H)/\gamma_{\rm N}$, 
as a function of reduced magnetic fields, $H/H_{\rm c2}$, 
for (a) Y(Ni$_{1-x}$Pt$_{x}$)$_{2}$B$_{2}$C and (b) Nb$_{1-x}$Ta$_{x}$Se$_{2}$. 
The solid lines are guides to the eye. The insets show 
the effective vortex core radius, $\rho_{\rm V}$, determined by assuming 
$2\pi{\rho_{\rm V}(H)}^{2}\gamma_{\rm N} = \gamma(H)/(H/\Phi_{0})$, 
as a function of $H/H_{\rm c2}$. 
$\rho_{\rm V}$ of NbSe$_{2}$ determined by STS
~\protect\cite{Golubov} 
at $T/T_{\rm c}$ = 0.33 
are shown (open squares) for comparison. 
}
\label{Fig3}
\end{figure}
%----------------------------------------------------------------------

%-------------------------------Fig. 4---------------------------------
\begin{figure}
\begin{center}
\leavevmode
\epsfxsize=250pt    % set the width
\epsfbox{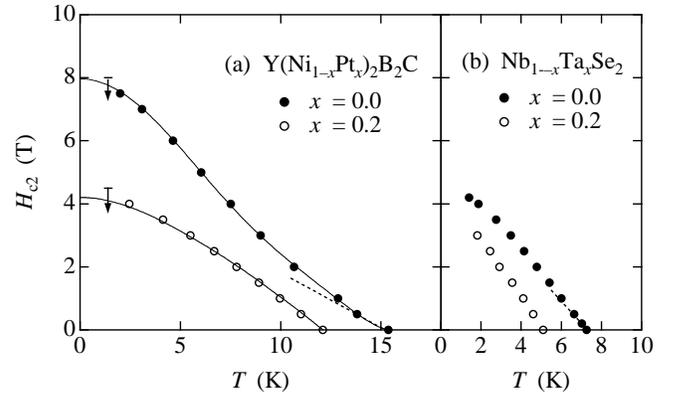}
\end{center}
\caption{
Temperature dependence of the upper critical field $H_{\rm c2}(T)$ for
Y(Ni$_{1-x}$Pt$_{x}$)$_{2}$B$_{2}$C and Nb$_{1-x}$Ta$_{x}$Se$_{2}$. The solid 
lines are guides to the eye. The broken 
lines indicates the fit of the initial slope of $H_{\rm c2}(T)$ at $T_{\rm c}$. 
}
\label{Fig4}
\end{figure}
%----------------------------------------------------------------------

\section{Table}

%--------------------------------Table-------------------------------
\begin{table}
\caption{
Superconducting and normal state parameters for the 
Y(Ni$_{1-x}$Pt$_{x}$)$_{2}$B$_{2}$C and 
Nb$_{1-x}$Ta$_{x}$Se$_{2}$ samples studied. 
The mean free path $l$ was calculated from the resistivity $\rho_{0}$ 
just above $T_{\rm c}$ by using 
$l=\hbar{(3{\pi}^{2})}^{1/3}/{e}^{2}{n}^{2/3}{\rho}_{0}$. 
We used  an electron density of $3\times 10^{22}$ {\it e}/cm$^{3}$ 
for Y(Ni$_{1-x}$Pt$_{x}$)$_{2}$B$_{2}$C and 
$1.6\times 10^{22}$ {\it e}/cm$^{3}$ for Nb$_{1-x}$Ta$_{x}$Se$_{2}$. 
The Ginzburg Landau coherence length parallel to the $ab$ plane, 
$\xi(0)$, was determined from $H_{\rm c2}(0)$. 
}
\label{Table1}
\begin{tabular}{ccccccc}
\hline
    & $x$ & $T_{\rm c}$ & $\rho_{0}$ & RRR & $\xi(0)$ & $l$ \\
    &     &  (K)    & ($\mu\Omega$cm) &   &  (\AA) & (\AA) \\
\hline
 Y(Ni$_{1-x}$Pt$_{x}$)$_{2}$B$_{2}$C & 0.0 & 15.4 & 0.87 & 37.4 & 65 & 1500 \\
   & 0.2 & 12.1 & 35.5 &  2.6 & 90 &   38 \\
 Nb$_{1-x}$Ta$_{x}$Se$_{2}$ & 0.0 &  7.3 &  4.7 & 20.0 & 85 &  450 \\
   & 0.2 &  5.1 & 25.4 &  4.2 &100 &   80 \\
\hline
\end{tabular}
\end{table}
%----------------------------------------------------------------------

\end{document}